\documentclass[12pt,a4paper]{article}
\usepackage{graphicx}
\usepackage{times}
\title{\bf Observational approach to the chemical evolution of high-mass binaries}
\author{K. Pavlovski$^{1,2}$, J. Southworth$^2$, E. Tamajo$^1$, and V. Kolbas$^1$\\
\vspace{1cm}\\
\normalsize $^1$ Department of Physics, University of Zagreb, Croatia \\
\normalsize $^2$ Astrophysics Group, Keele University, Staffordshire, UK}
\date{\mbox{}}
\begin{document}
\maketitle
\pagestyle{empty}
%
%
\def\bull{\vrule height .9ex width .8ex depth -.1ex}
\makeatletter
\def\ps@plain{\let\@mkboth\gobbletwo
\def\@oddhead{}\def\@oddfoot{\hfil\tiny\bull\quad
``The multi-wavelength view of hot, massive stars''; 39$^{\rm th}$ Li\`ege Int.\ Astroph.\ Coll., 12-16 July 2010 \quad\bull}%
\def\@evenhead{}\let\@evenfoot\@oddfoot}
\makeatother
%
%
\def\beginrefer{\section*{References}%
\begin{quotation}\mbox{}\par}
\def\refer#1\par{{\setlength{\parindent}{-\leftmargin}\indent#1\par}}
\def\endrefer{\end{quotation}}
%
%
{\noindent\small{\bf Abstract:}
The complexity of composite spectra of close binaries makes the study of the individual stellar
spectra extremely difficult. For this reason there exists very little information on the chemical
composition of high-mass stars in close binaries, despite its importance for understanding
the evolution of massive stars and close binary systems. A way around this problem exists:
spectral disentangling allows a time-series of composite spectra to be decomposed into
their individual components whilst preserving the total signal-to-noise ratio in the input
spectra. Here we present the results of our ongoing project to obtain the atmospheric
parameters of high-mass components in binary and multiple systems using spectral disentangling.
So far, we have performed detailed abundance studies for 14 stars in eight eclipsing binary systems.
Of these, V380\,Cyg, V\,621 Per and V453\,Cyg are the most informative as their primary components are
 evolved either close to or beyond the TAMS. Contrary to theoretical predictions of rotating
single-star evolutionary models, both of these stars show no abundance changes relative
to unevolved main sequence stars of the same mass. It is obvious that other effects are
important in the chemical evolution of components in binary stars. Analyses are ongoing
for further systems, including AH\,Cep, CW\,Cep and V478\,Cyg.}
%
%

\section{Introduction}

In the last decade theoretical stellar evolutionary models, particularly for higher masses, were
improved considerably with the inclusion of additional physical effects beyond the standard ingredients.
 It was found that rotationally induced mixing and magnetic fields could cause substantial changes
in the resulting predictions (Meynet \& Maeder 2000, Heger \& Langer 2000). Some of these concern
 evolutionary changes in the chemical composition of stellar atmospheres. Due to the CNO cycle
in the core of high-mass stars some elements are enhanced (such as helium and nitrogen), some
are depleted (e.g.\ carbon), whilst some (e.g.\ oxygen) are not affected at all. The rotational
 mixing predicted by stellar models is so efficient that changes in the atmospheric composition
should be identifiable whilst the star is still on the main sequence.

On the observational side, substantial progress has also been made. The VLT/FLAMES survey
(Evans et al.\ 2005) produced CNO abundances for a large sample of B stars in the Milky Way,
and the Magellanic Clouds. This survey has opened new questions since a large population
of slow rotators have shown an enhancement of nitrogen (Hunter et al.\ 2009).
Also, important empirical constraints on models arose from the observational study
performed by Morel et al.\ (2008) who found that magnetic fields have an important
effect on the atmospheric composition of these stars.

\begin{figure}[h]
\centering
\includegraphics[width=14.5cm]{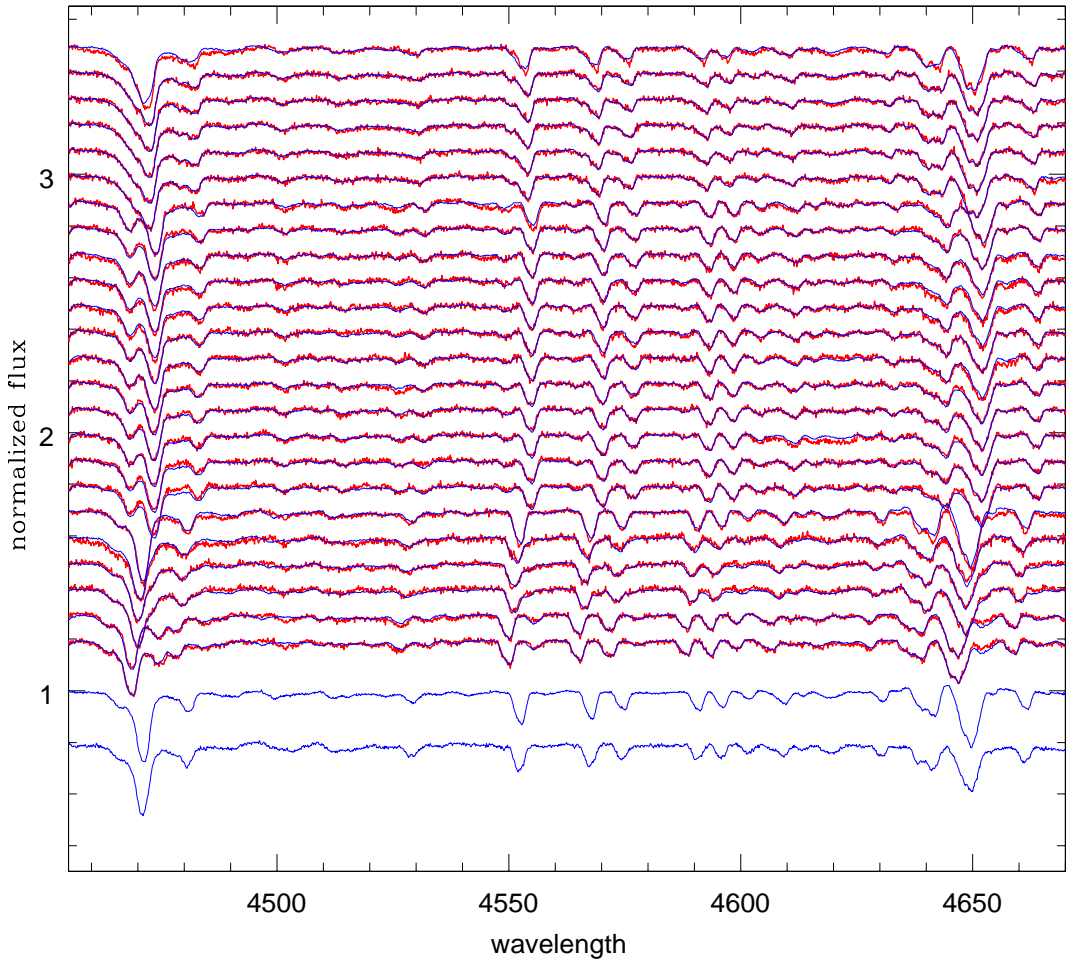}
\caption{Time series of observed composite spectra (red lines) of the B0\,V + B1\,V
close eclpsing binary system V453\,Cyg (Pavlovski \& Southworth 2009). This is a portion of \'echelle
spectra secured with the FIES spectrograph at the Nordic Optical Telescope (La Palma).
The individual disentangled spectra of the two stars, which have very similar effective
temperatures, are shown at the bottom of the plot (blue lines, secondary offset by $-0.2$)
with their correct continuum levels. The disentangled spectra have been adjusted with the
appropriate Doppler shifts and relative intensities to reproduce the observed spectra,
and are overlaid on them using blue lines.} \end{figure}


Detached eclipsing binaries are fundamental objects for obtaining empirical constraints
on the structure and evolution of high-mass stars, and are the primary source of directly
 measured stellar properties. Accurate physical properties are available for fewer than
a dozen high-mass binaries, and most have no observational constraints on their chemical
 composition (Torres, Andersen \& Gim\'enez 2010). The aim of our projects is to obtain
a sample of high-mass binaries both with accurate parameters and, for the first time,
with detailed abundance studies of the individual stars. We aim to gain insight into the chemical
evolution of high-mass stars in close binary systems. The close proximity of the components
leads to strong tidal forces, which may be an important additional effect on the internal
and chemical structure of the stars, beside rotation and magnetic fields.

\section{Sample and Method}

The complexity of the composite spectra of close binaries makes studying the spectra of
the individual stars extremely difficult. For this reason there exists very little information
on the chemical composition of high-mass stars in close binaries, despite its importance
for understanding the evolution of both massive stars and close binaries. A way around this
problem exists: spectral disentangling. This technique allows a time-series of composite
spectra to be decomposed into their individual components whilst preserving the total
signal-to-noise ratio in the input spectra, and without the use of template spectra
(Simon \& Sturm 1994). An overview of almost a dozen methods for spectral disentangling
 has been given by Pavlovski \& Hensberge (2010). For our work we use the {\sc fdbinary}
Fourier-space code (Iliji\'c et al.\ 2004). Synthetic spectra are generated using {\sc atlas9}
with non-LTE model atoms (see Pavlovski \& Southworth 2009 and Pavlovski et al.\ 2009 for details).

\begin{figure}[h]
\centering
\includegraphics[width=14.2cm]{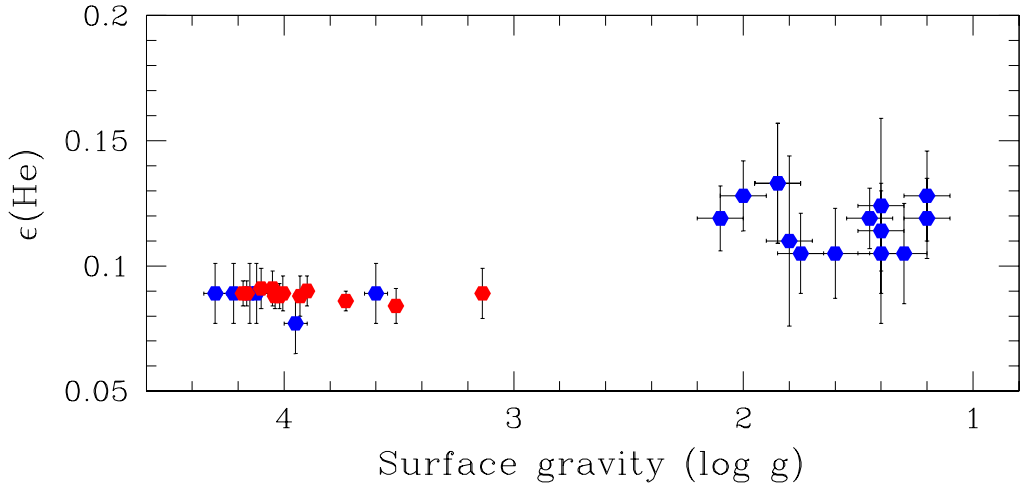}
\caption{Helium abundances for high-mass stars in close binaries from our sample (red symbols) compared to
 single sharp-lined B-type main sequence stars and BA supergiants in the Przybilla et al.\ (2010) sample (blue
symbols). $\epsilon$(He) is the fractional helium abundance by number of atoms.}
\end{figure}

A vital step in a spectroscopic abundance study is precise determination of the stellar atmospheric
parameters (effective temperature, surface gravity, microturbulence velocity, etc). When reconstructing
 the separate spectra of the components their individual light contributions have to be obtained either
 from the disentangled spectra itself, or from some other source such as a complementary light curve analysis
(c.f.\ Pavlovski \& Hensberge 2010).

So far, we have performed detailed abundance studies for 14 components in eight eclipsing binaries.
In many cases we have also reanalysed existing or new light curves. Of the systems studied,
 V380\,Cyg (Pavlovski et al.\ 2009), V453\,Cyg (Pavlovski \& Southworth 2009, Southworth et al.\ 2004a)
and V621\,Per (Southworth et al., 2004b, 2011 in prep.) are the most informative as their primary
components are evolved either close to or beyond the terminal-age main sequence (TAMS). Other binaries
studied include V578\,Mon (Pavlovski \& Hensberge 2005, see also Hensberge et al.\ 2000),
AH\,Cep, CW\,Cep, Y\,Cyg and V478\,Cyg [helium abundances have also been measured from disentangled spectra
for DH\,Cep (Sturm \& Simon 1994), Y\,Cyg (Simon, Sturm \& Fiedler 1994) and DW\,Car (Southworth \& Clausen 2007)].
These objects mostly contain stars at the beginning of their main sequence lifetimes, so are important
for calibrating theoretical models near their initial conditions.

\section{The quest for surface helium enrichment}  

Theoretical stellar evolutionary models which include rotational mixing predict an enrichment of helium at
the stellar surface, even during a star's main sequence lifetime. Extensive observational studies
comprising B-type stars in the field (Lyubimkov et al.\ 2004), and in stellar clusters (Huang \& Gies 2006)
yield evidence for this enrichment, but with a very large scatter in the individual measurements.
An unexpectedly large fraction of both helium-rich, and helium-weak stars were detected by Huang \& Gies (2006),
who included only three helium lines in their analysis.

The results of our detailed abundance determinations in the sample of 14 components of close binary
stars are shown in  Fig.\ 2 (red symbols). The results of a recent study of helium abundances in
the sample of sharp-lined main sequence and BA supergiants (Przybilla et al.\ 2010) are also plotted (blue symbols).
It is interesting that no helium abundance enrichment has been detected in these studies,
either for single stars (Przybilla et al.\ 2010) or the components of close binaries (this work).
The studies therefore do not support a large spread in helium abundance, as found by other authors,
with the caveat that the sample of main sequence stars studied is limited.

\begin{figure}[h]
\centering
\includegraphics[width=15cm]{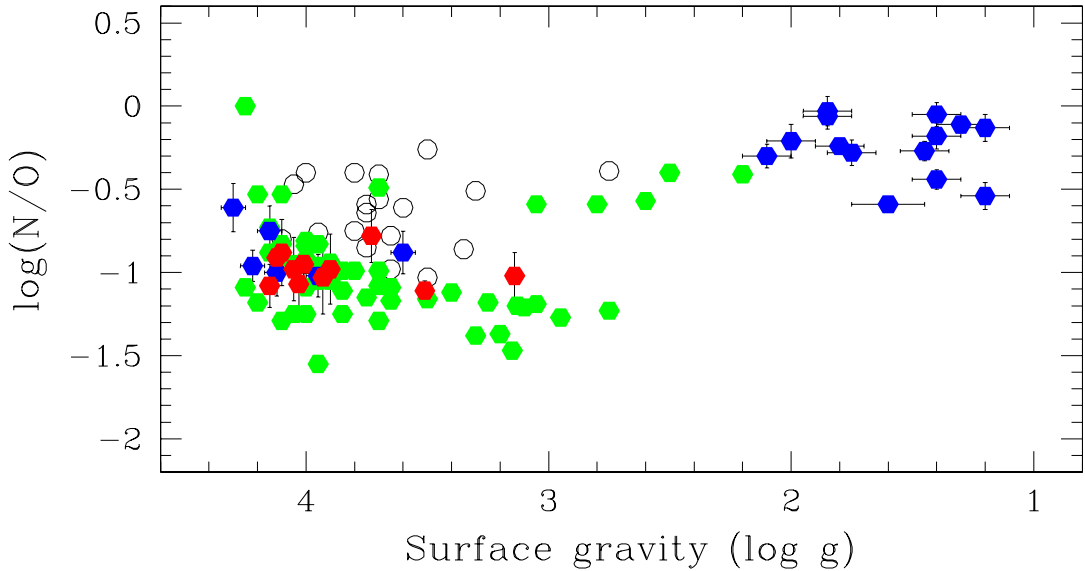}
\caption{Evolution of nitrogen in high-mass MS stars to supergiants.
The close binary systems in our sample are represented by red symbols. Other
symbols represent single stars as follows: blue symbols the VLT/FLAMES survey of B stars in
Milky Way (Hunter et al.\ 2009); green symbols the results of an abundance study for a sample
of B stars with detected magnetic fields (Morel et al.\ 2008); and open symbols a study of
sharp-lined stars (Przybilla et al.\ 2010).}
\end{figure}

\section{The evolution of nitrogen in high-mass binaries}

In close binary stars, both fast rotation and tidal forces due to the proximity of the components play
an important role in stellar evolution. Tides spin up (or spin down) the stars until their rotation
period synchronises with the orbital period. The effects of tides, rotational mixing and magnetic
fields were studied by de Mink et al.\ (2009). Their model calculations indicate a significant
dependence of the surface helium and nitrogen abundances for short-period systems ($P < 2$ days)
for a considerable fraction of their MS lifetime. The best candidates for testing these concepts
contain more massive components, in advanced phases of the core hydrogen-burning phase, with
significantly less massive and less evolved companions. V380\,Cyg, V621\,Per and V453\,Cyg fit
this bill well, but have longer orbital periods (3.9\,d to 25.5\,d) so are not predicted to show significant
 abundance enhancements.
This is illustrated in Fig.\,3 in which the abundance ratio N/O is plotted against $\log g$, which is a
good indicator of evolutionary stage. The N/O ratios for the evolved stars in our sample are consistent
with ZAMS values, like many of the stars in the VLT/FLAMES sample of Hunter et al.\ (2009).
The evolutionary enhancement of nitrogen is only clearly present in the sample of supergiants observed by Przybilla et al.\ (2010).
On average the magnetic B-type stars (Morel et al.\ 2008) have large nitrogen abundances, but definitive conclusions
on the role of magnetic fields on nitrogen enrichment are still not possible (Morel 2010). The large spread in
nitrogen abundances for MS stars is obvious.

Since the enhancements of helium and nitrogen are larger at lower metallicity,
the best candidates for detailed study would be close binaries in the Magellanic Clouds. However,
these are challenging objects for accurate abundance determination due to their high rotational velocities (resulting in line blending) and relative faintness.

\section*{Acknowledgements}

KP acknowledges receipt of the Leverhulme Trust Visiting Professorship which enables him to work at
 Keele University, UK, where this work was performed. This research is supported in part by a grant
to KP from the Croatian Ministry of Science and Education.

%
%
\footnotesize
\beginrefer
\refer De Mink, S., Cantiello, M., Langer, N., Pols, O.R., Brott, I., Yoon, S.-Ch., 2009, A\&A, 497, 243

\refer Evans, C.J., Smartt, S.J., Lee, J.-K., et al., 2005, A\&A, 437, 467

\refer Heger, A., Langer, N., 2000, ApJ, 544, 1016

\refer Hensberge, H., Pavlovski, K., Verschueren, W., 2000, A\&A, 358, 553

\refer Huang, W., Gies, D.R., 2006, ApJ, 648, 591

\refer Hunter, I., Brott, I., Langer, N., et al., 2009, A\&A, 496, 841

\refer Iliji\'c S., Hensberge H., Pavlovski K., Freyhammer L.M., 2004, in Hilditch
R.W., Hensberge H., Pavlovski K., eds, ASP Conf. Ser. Vol. 318, {\it Spectroscopically
and Spatially Resolving the Components of Close Binary
Systems}. Astron. Soc. Pac., San Francisco, p.\ 111

\refer Lyubimkov, L.S., Rostophchin, S.I., Lambert, D.L., 2004, MNRAS, 351, 745

\refer Meynet, G., Maeder, A., 2000, A\&A, 361, 101

\refer Morel, T., 2010, this proceedings ({\sf arXiv:1009.3433})

\refer Morel, T., Hubrig, S., Briquet, M., 2008, 481, 453

\refer Pavlovski, K., Hensberge, H., 2005, A\&A, 439, 309

\refer Pavlovski, K., Hensberge, H., 2010, in {\it Binaries - Key to Comprehension of the
Universe}, eds.\ A.\ Pr\v{s}a and M.\ Zejda,  ASP Conference Series (in press); arXiv:0909.3246

\refer Pavlovski, K., Southworth, J., 2009, MNRAS, 394, 1519

\refer Pavlovski, K., Tamajo, E., Koubsky, P., Southworth, J., Yang, S.,  Kolbas, V., 2009, MNRAS, 400, 791

\refer Przybilla, N., Firnstein, M., Nieva, M.F., Meynet, G., Maeder, A., 2010, A\&A, 517, 38

\refer Simon, K.P., Sturm, E., 1994, A\&A, 281, 286

\refer Simon, K.P., Sturm, E., Fiedler, A., 1994, A\&A, 292, 507

\refer Southworth, J., Clausen J.\ V., 2007, A\&A, 461, 1077

\refer Southworth, J., Maxted, P.F.L., Smalley, B., 2004a, MNRAS, 351, 1277

\refer Southworth, J., Zucker, S., Maxted, P.F.L., Smalley, B., 2004b, MNRAS, 355, 986

\refer Sturm, E., Simon, K.P., 1994, A\&A, 282, 93

\refer Torres, G., Andersen, J., Gim\'{e}nez, A., 2010, ARA\&A, 18, 67

\endrefer
\end{document}